\magnification=1200

\pageno=1
\centerline {\bf SU($\infty$) $q$-MOYAL-NAHM EQUATIONS AND QUANTUM }
\centerline {\bf DEFORMATIONS  OF THE SELF DUAL MEMBRANE }
\medskip
\centerline {\bf Carlos Castro }
\centerline {\bf Center for Particle Theory, Physics Dept.}
\centerline {\bf University of Texas}
\centerline {\bf Austin , Texas 78712}

\smallskip
\centerline {\bf March , 1997} 
\smallskip
\centerline {\bf ABSTRACT}

Since the lightcone  self dual spherical membrane,  moving in flat target backgrounds,  has a direct correspondence with the $SU(\infty)$ Nahm equations and 
the continuous Toda theory, we construct the quantum/Moyal deformations of the self dual membrane in terms of the $q$-Moyal star product . 
The  $q$ deformations of the $SU(\infty)$ Nahm equations are studied and explicit solutions are given. The continuum limit of the $q$ Toda chain equations are obtained furnishing $q$ deformations of the self dual membrane.  Finally, the continuum  Moyal-Toda chain equation is embedded into the $SU(\infty)$ Moyal-Nahm equations, rendering the relation with the Moyal deformations of the self dual membrane. $W_{\infty}$ and $q$-$W_{\infty}$ algebras arise as the symmetry algebras and the role  of (  the recently developed ) $quantum$ Lie algebras associated with quantized universal enveloping algebras is pointed out  pertaining the formulation of a $q$ Toda theory. We review as well the     
Weyl-Wigner-Moyal quantization of the 3D continuous Toda field equation, and its associated 2D continuous Toda molecule, based on Moyal deformations of rotational Killing symmetry reductions of Plebanski first heavenly equation.


\smallskip
\centerline {\bf I. Introduction}

The quantization program of the $3D$ continuous Toda theory ( $2D$ Toda molecule) is a challenging enterprise that we believe would enable  to understand the quantum dynamics and spectra of the quantum self dual membrane [1]. The classical theory can be 
obtained from a rotational Killing symmetry reduction of the $4D$ Self Dual 
Gravitational (SDG)  equations expressed in terms of Plebanski first heavenly form that furnish ( complexified) self dual metrics of the form :
$ds^2 =\partial_{x^i}\partial_{{\tilde x}^j}\Omega dx^i d{\tilde x}^j.$ for $x^i=y,z$; 
${\tilde x}^j ={\tilde y},{\tilde z}$ and $\Omega$ is Plebanski first heavenly form. The latter equations can, in turn, be obtained from a dimensional reduction of the $4D~SU(\infty)$ Self Dual Yang Mills equations (SDYM), an effective $6D$ theory [4,5] and references therein. The Lie algebra $su(\infty)$ was shown to be isomorphic ( in a basis dependent limit) to the Lie algebra of area preserving 
diffeomorphisms of a $2D$ surface, $sdiff~(\Sigma)$ [6]. 

It was pointed out in [1] that a Killing symmetry reduction of the  
$4D$ Quantized Self Dual Gravity ,  via the $W_\infty$ co-adjoint orbit method [7,8] , gives a quantized Toda theory. In this letter we shall present a more direct quantization method and quantize the Toda theory using the Weyl-Wigner-Moyal prescription (WWM). A WWM description of the $SU(\infty)$ Nahm equations was carried out by [9] and a correspondence between BPS magnetic monopoles and 
hyper Kahler metrics was provided. There is a one-to-one correspondence between solutions of the Bogomolny equations with appropriate boundary conditions and solutions of the $SU(2)$ Nahm equations.

As emphasized in [9], BPS monopoles are solutions of the Bogomolny equations whose role has been very relevant in the study of $D~3$-branes realizations of $N=2~D=4$ super YM theories in IIB superstrings [10]; $D$ instantons constructions [11]; in the study of moduli spaces of BPS monopoles and origins of ``mirror''  symmetry 
in $3D$ [12]; in constructions of self dual metrics associated with hyper Kahler spaces [13,14], among others. 

Using our results of [15] based on [9,16] we show in {\bf II} that a WWM [17] quantization approach yields a straighforward quantization scheme for the $3D$ continuous 
Toda theory ( $2D$ Toda molecule). Supersymmetric extensions can be carried out following [4] where we wrote down the supersymmetric analog of Plebanski equations for SD Supergravity. Simple solutions are proposed. 

There are fundamental differences between our results and those which have appeared in the literature [9]. Amongst these are  (i) One is $not$ taking the limit of $\hbar \rightarrow 0$ while having $N=\infty$ in the classical $SU(N)$ Nahm equations.   
(ii) We are working with the $SU(\infty)$ Moyal-Nahm equations and not with the $SU(2)$ Moyal-Nahm equations. Hence, we have $\hbar \not= 0;N=\infty$ simultaneously. (iii) The connection with the self dual membrane and $W_{\infty}$ algebras was proposed in [1] by the author.   
The results of [9] become very useful in the implementation of the WWM quantization program and in the embedding of the $SU(2)$ Moyal-Nahm solutions [9] into the $SU(\infty)$ Moyal-Nahm equations studied in the present work. 

In {\bf III } we construct a continuum $q$ Toda theory using the $q$-star product developed in 
[25]. Explicit solutions to the $SU(\infty)$ $q$-Nahm equations are found 
and the continuum limit of the $q$ Toda molecule equation is defined based on the recently developed concept of $quantum$ Lie algebras [30] . The relation to the $q$ deformations of the self dual membrane is then furnished. Finally in {\bf IV} we discuss the $SU(\infty)$ Moyal-Nahm system in relation to the Moyal deformations of the self dual membrane.

The study of $quantum$ Lie algebras [30] grew out of the desire to understand 
exact results obtained in quantum affine Toda theories. It has been possible to obtain full quantum mass ratios and exact $S$ matrices for the fundamental 
particles [43,44,45]. In the quantum theory mass receive corrections; however, exact quantum masses are still given by a formula involving the quantum twisted Coxeter number of the algebra $g$. Furthermore, the existence of the solutions of the locations of the poles of the $S$ matrix can be understood in terms of properties of $quantum$ root systems of $quantum$ Lie algebras [30]. These algebras are discussed in {\bf III} before the $q$ continuum Toda is defined in terms  of the quantum Lie algebra denoted by ${\cal L}_q (su(\infty)$. 

We expect that a $q$-Moyal deformation program of the self dual membrane might 
yield important information about how to quantize  the full quantum 
membrane theory beyond the self dual exactly integrable sector.

\centerline {\bf II The Moyal Quantization of the Continuous Toda Theory}

In this section we shall present the Moyal quantization of the continuous Toda theory. The Moyal deformations of the rotational Killing symmetry reduction of Plebanski self dual gravity equations in $4D$ were given by the author in [15] based on the results of [16]. Starting with  :

$$\Omega (y,{\tilde y},z,{\tilde z};\kappa )\equiv 
\sum_{n=0}^\infty (\kappa/{\tilde y})^n \Omega_n (r,z,{\tilde z}). \eqno (1)$$
where each $\Omega_n$ is only a function of the complexified variables $r\equiv y{\tilde y}$ and $z,{\tilde z}$ . Our notation is the same  from [15]. A real slice may be taken by setting ${\tilde y} ={\bar y}, {\tilde z}={\bar z},..$.   
The Moyal deformations of Plebanski's equation read  :

$$\{\Omega_z, \Omega_{{y}} \}_{Moyal} =1. \eqno (2)$$
where the Moyal bracket is taken w.r.t the  ${\tilde z}, {\tilde y}$ 
variables. In general, the Moyal  bracket may  defined as a power expansion in the deformation parameter, $\kappa$ :

$$\{f,g\}_{{\tilde y},{\tilde z}} \equiv [\kappa^{-1} sin~\kappa (\partial_{{\tilde y}_f}
\partial_{{\tilde z}_g} -  \partial_{{\tilde y}_g}
\partial_{{\tilde z}_f})]fg. \eqno (3)$$
with the subscripts under ${\tilde y}, {\tilde z}$ denote derivatives acting only on $f$ or on $g$ accordingly.

We begin by writing down the derivatives w.r.t the $y,{\tilde y}$ variables 
when these are acting on $\Omega$

$$\partial_{y}={1\over y}r\partial_r.~        
\partial_{{\tilde y}}={1\over {\tilde y}}r\partial_r.   \eqno (4)$$
$$\partial_y \partial _{{\tilde y}}=r\partial^2_r +\partial_r.~
\partial^2_{{\tilde y}}=({1\over {\tilde y}})^2 (r^2\partial^2_r +r\partial r).
$$

$$\partial^3_{{\tilde y}}=({1\over {\tilde y}})^3 (
r^3\partial^3_r +r^2\partial^2_r -r\partial r).
$$
$$...................................$$

Hence, the Moyal bracket (2)
yields the infinite number of equations after matching, order by order in $n$, powers of $(\kappa/ {\tilde y})$:

$$\{\Omega_{0z}, \Omega_{0{y}} \}_{Poisson} =1\Rightarrow (r\Omega_{0r})_r
\Omega_{0z{\tilde z}}-r\Omega_{0rz}\Omega_{0r{\tilde z}}=1. \eqno (5)$$
$$\Omega_{0z{\tilde z}}[-\Omega_{1r}+(r\Omega_{1r})_r]-
r\Omega_{1r{\tilde z}}
\Omega_{0rz}+\Omega_{1z{\tilde z}}(r\Omega_{0r})_r 
+\Omega_{0r{\tilde z}}(\Omega_{1z}-r\Omega_{1rz})=0. $$
$$......................$$
$$
.......................$$
the subscripts represent partial derivatives of the functions 
$\Omega_n (r=y{\tilde y},z,{\tilde z})$ for $ n=0,1,2.....$ w.r.t the variables   $r,z,{\tilde z}$ in accordance with the Killing symmetry reduction conditions. The first equation, after a nontrivial change of variables, can be recast as the $sl(\infty)$ continual Toda equation as demonstrated [2,3] . The remaining equations are the Moyal 
deformations. The symmetry algebra of these equations is the Moyal deformation of the classical $w_\infty$ algebra which turns out to be precisely the centerless $W_\infty$ algebra as shown by [19]. Central extensions can be added using the cocycle formula in terms of logarithms of derivative operators [20] giving the $W_\infty$ algebra first built by [21]. .

From now on in order not to be confused with the notation of [5] we shall denote for ${\tilde \Omega}(y',{\tilde y}',z',{\tilde z}';\kappa)$ to be the solutions to eq-(2). The authors [5] used $\Omega (z+{\tilde y},{\tilde z}-y,q,p;\hbar)$ as solutions to the Moyal deformations of Plebanski equation. The dictionary from the results of [15], given by eqs-(1-5), to the ones  used by the authors of [5] is 
obtained  from the relation :

$$\{ {\tilde \Omega}_{z'}, {\tilde \Omega}_{y'}\}_{{\tilde z}', {\tilde y}'}
 =\{ \Omega_w, \Omega_{{\tilde w}}\}_{{q,p}}=1.~\kappa=\hbar.~w=z+{\tilde y}.~
{\tilde w}={\tilde z}-y. \eqno (6a)$$
For example,  the four conditions : ${\tilde \Omega}_{z'}=\Omega_w;
  {\tilde \Omega}_{y'}=\Omega_{{\tilde w}}$ and ${\tilde z}'=q;{\tilde y}'=p$
are one of many which satisfy the previous dictionary relation (6a). One could perform a deformed-canonical transformation from ${\tilde z}',{\tilde y}'$ to the new variables $q,p$ iff the Moyal bracket $\{q,p\}=1$. Clearly, the simplest canonical transformation is the one chosen above. 
The latter four conditions yield the transformation rules from 
${\tilde \Omega}$ 
to  $\Omega$. The change of coordinates :
$${\tilde z}'=q.~{\tilde y}'=p.~z'=z'(w,{\tilde w},q,p|\Omega).~y'=y'(w,{\tilde w},
q,p|\Omega).\eqno (6b)$$
leads to :
$$z'=w+f(p,q).~y'={\tilde w}+g(p,q).$$
once one sets :
$${\tilde \Omega}[z'(w,{\tilde w}....);y'(w,{\tilde w}...,);{\tilde z}'=q;{\tilde y}'=p]=\Omega (w,{\tilde w},q,p). \eqno (6c)$$
for ${\tilde \Omega},\Omega$ obeying eqs-(2,6a). The $implicitly$ defined change of coordinates by the four conditions stated above is clearly dependent on the family of solutions to eqs-(2,6a). It is highly nontrivial. The reason this is required is because the choice of variables must be consistent with those of [9] to implement the WWM formalism.  
For example, choosing $\Omega=\Omega_o =z'{\tilde z}'+y'{\tilde y}'$ as a solution to the eqs-(2,5) yields for (6b) :

$$z'=w+{\lambda \over q}.~y'={\tilde w} -{\lambda \over p}. \eqno (6d)$$
The reality conditions on $w,{\tilde w}$ may be chosen to be :
${\tilde w}={\bar w}$ which implies ${\tilde z} ={\bar z};{\tilde y}=-{\bar y}$. It differs from the reality condition chosen for the original variables. It is important to remark as well that the variables $p,q$ are also complexified and the  area-preserving algebra is also : the algebra is $su^*(\infty)$ [4].

Now we can make contact with the results of [5,9]. 
In general, the expressions  that relate  the $6D$ scalar field 
$\Theta (z,{\tilde z},y,{\tilde y},q,p;\hbar)$ to the 
$4D~SU(\infty)$ YM potentials become, as a result of the dimensional reduction of the effective $6D$ theory to the $4D$ SDG one,  the following  [4,5] :

$$\partial_z \Theta =\partial_{{\tilde y}}\Theta =\partial_{{w}}\Theta.         ~~~\partial_y \Theta =-\partial_{{\tilde z}}\Theta=
-\partial_{{\tilde w}} \Theta. \eqno (7a)$$
with $\kappa \equiv \hbar$ and $w=z+{\tilde y};{\tilde w}={\tilde z} -y$. Eqs-(7a) are basically equivalent to the integrated dimensional reduction condition : 

$$\Theta (z,{\tilde z},y,{\tilde y},q,p;\hbar)=\Omega (z+{\tilde y},{\tilde z}-y,q,p;\hbar )\equiv \Omega (w,{\tilde w},q,p;\hbar)\equiv \sum_{n=0}^\infty (\hbar)^n \Omega_n (r=w{\tilde w};q,p). \eqno(7b)$$
which furnishes the Moyal-deformed YM potentials : 
$$A_{{\tilde z}} ({\tilde y},w,{\tilde w} ,q,{p};\hbar)=
\partial_{{\tilde w}} 
\Omega (w,{\tilde w},q,p;\hbar) +{1\over 2}{\tilde y}.~ 
A_{{\tilde y}} ({\tilde z},w, {\tilde w} ,q,{p};\hbar)  =\partial_w \Omega (w,\tilde w,q,p;\hbar ) -{1\over 2}{\tilde z}.\eqno (8)$$. 
One defines the linear combination of the YM potentials  :

 $$A_{{\tilde z}}-A_{y}=A_{{\tilde w}}.
~A_{{\tilde y}}+A_{z}=A_{w} \eqno (9)$$
The new fields are denoted by $A_{w}, A_{{\tilde w}}$. After the following 
gauge conditions are chosen  
$A_z=0, A_y=0,$ [5] ,  it follows that $A_{{\tilde z}}=A_{{\tilde w}}$ and 
$A_{{\tilde y}}=A_{{w}}$.

For every solution of the infinite number of eqs-(5) by succesive iterations,  one has the $corresponding$ 
solution for the YM potentials given by eqs-(8) that are $associated$ with the 
Moyal deformations of the Killing symmetry $reductions$ of Plebanski first 
heavenly equation. Therefore,  YM potentials obtained from (5) and (8) $encode$ the 
Killing symmetry reduction. In eq-(14) we shall see that the operator equations of motion   corresponding to the Moyal quantization process of the Toda theory involves solely the operator ${\hat \Omega}$.  However, matters are not that simple because to solve the infinite number of equations (5) iteratively is far from trivial.  
The important fact is that in principle one has a systematic way of 
$solving$ (2).

The authors [9] constructed solutions to the Moyal deformations of the 
$SU(2)/SL(2)$ Nahm's equations employing the Weyl-Wigner-Moyal (WWM)  map which required the use of  
$known$ representations of $SU(2)/SL(2)$ Lie algebras [22]  in terms of 
 operators acting in the Hilbert space, $L^2(R^1)$. Also known in [9] were the solutions to the classical $SU(2)/SL(2)$ Nahm equations in terms of elliptic 
functions. The ``classical'' $\hbar \rightarrow 0$ limit of the WWM quantization of the $SU(2)$ Nahm equations was equivalent to the $N\rightarrow \infty$ limit of the 
$classical$ $SU(N)$ Nahm equations and, in this fashion, hyper Kahler metrics 
of the type discussed by [13,14] were obtained.

Another important conclusion that can be inferred from [5,9] is that one can embed the WWM-quantized $SU(2)$ 
solutions of the Moyal-deformed $SU(2)$ Nahm equations found in [9] into the $SU(\infty)$ Moyal-deformed Nahm equations and have, in this way, exact quantum solutions to the Moyal deformations of the $2D$ continuous Toda molecule which was essential in the construction of the quantum self dual membrane [1]. Since a dimensional reduction of the $W_\infty \oplus {\hat W}_\infty$ algebra is the symmetry algebra of the $2D$ effective theory, algebra that was coined $U_\infty$ in [1], one can generate other quantum solutions by $U_\infty$ co-adjoint orbit actions of the special solution found by [9]. One has then recovered the Killing symmetry reductions of the Quantum $4D$ Self Dual Gravity via the $W_\infty$ co-adjoint 
orbit method [7,8].   

The case displayed here is the $converse$. We do not have ( as far as we know) $SU(\infty)$ representations in $L^2(R^1)$. However, we can in principle $solve$ (5) iteratively. The goal is now to retrieve the operator coresponding to $\Omega (w,{\tilde w},q,p;\hbar)$.

The WWM formalism [17]  establishes the one-to-one map 
that takes self-adjoint operator-valued quantities, 
${\hat \Omega}(w,{\tilde w})$, living  on the $2D$ space parametrized by  coordinates, $w,{\tilde w}$,  and acting in  the Hilbert space of $L^2 (R^1)$,   
to the space of smooth functions on the phase space manifold
${\cal M}(q,p)$ associated withe real line, $R^1$. 
The map is defined :

$$ \Omega (w,{\tilde w},q,p;\hbar)
\equiv \int^\infty_{-\infty}d\xi <q-{\xi\over 2}|{\hat \Omega } (w,{\tilde w})|q+{\xi \over 2}>exp[{i\xi p\over \hbar}]. \eqno (10a)$$

Since the l.h.s of (10a) is completely determined in terms of solutions to 
eq-(2) after the iteration process in (5) and the use of the relation (6), the r.h.s is also known : the inverse transform yields the expectation values of the operator : 

$$  <q-{\xi\over 2}|{\hat \Omega } (w,{\tilde w})|q+{\xi \over 2}>=
\int^\infty_{-\infty}dp ~\Omega (w,{\tilde w},q,p;\hbar)
exp[-{i\xi p\over \hbar}]. 
\eqno (10b)$$
i.e. $all$ the matrix elements of the operator ${\hat \Omega}(w,{\tilde w})$ are determined from (10b), therefore the operator ${\hat \Omega}$ can be retrieved completely. The latter operator obeys the operator analog of the zero curvature condition, eq-(14), below. The authors in [23] have discussed ways to retrieve  distribution functions, in the quantum statistical treatment of photons, as expectation values of 
a density  operator in a diagonal basis of coherent states. Eq-(10b) suffices to obtain the full operator without the need to recur to the coherent ( overcomplete) basis of states.

It is well known by now that the SDYM equations can be obtained as a zero curvature condition [24].  In particular, eq-(2).  The operator valued extension of the zero-curvature condition reads :

$$\partial_{{\tilde z}} {\hat {\cal A}}_{{\tilde y}}-
\partial_{{\tilde y}}{\hat {\cal A}}_{{\tilde z}} +
{1\over i\hbar} [{\hat {\cal A}}_{{\tilde y}},
{\hat {\cal A}}_{{\tilde z}}]=0. \eqno (11)$$
which is the WWM transform of the original Moyal deformations of the zero curvature condition :
$$\partial_{{\tilde z}}A_{{\tilde y}}({\tilde z},q,p,w,{\tilde w};\hbar)-
\partial_{{\tilde y}}A_{{\tilde z}}({\tilde y},q,p,w,{\tilde w};\hbar) 
+\{ A _{{\tilde y}}
, A _{{\tilde z}}\}_{q,p}=0. \eqno (12)$$
This is possible due to the fact that the WWM formalism, the map ${\cal W}^{-1}$ 
preserves the Lie algebra commutation relation  :
$${\cal W}^{-1} ({1\over i\hbar}[{\hat {\cal O}}^i,{\hat {\cal O}}^j]) \equiv 
\{ {\cal O}^i,{\cal O}^j \}_{Moyal}. \eqno (13)$$

The latter equations (11,12)  can be recast $entirely$ in terms of $\Omega (w,{\tilde w},q,p,\hbar)$ and the operator ${\hat  \Omega} (w,{\tilde w})$ after one recurs to the relations $A_{{\tilde z}}=A_{{\tilde w}}; A_{{\tilde y}}=A_{w}$
(9) and the dimensional reduction conditions (7) :
$\partial_{{\tilde z}}=\partial_{{\tilde w}};
\partial_{{\tilde y}}=\partial_{ w}$. 
Hence, one arrives at the $main$ result of this section   :

$${1\over i\hbar}[{\hat \Omega}_w,{\hat  \Omega}_{{\tilde w}}]={\hat 1} \leftrightarrow 
\{ \Omega_w,{\Omega}_{{\tilde w}}\}_{Moyal}=1 . \eqno (14)$$
i.e. the operator ${\hat \Omega}$ obeys the operator equations of motion encoding the quantum dynamics. The carets denote operators. The operator form of eq-(14) was possible due to the fact that the first two terms in the zero curvature condition (12) are 
:
$$\partial_{{\tilde z}}A_{{\tilde y}}
-\partial_{{\tilde y}}A_{{\tilde z}}=-1. \eqno (15)$$
as one can verify by inspection from the dimensional reduction conditions in 
(7) and after using (8). 

The operator valued expression in (14) encodes the Moyal quantization of the 
continuous Toda field.  
The original continuous Toda equation  is [2,3,18]:  

$${\partial^2 \rho \over \partial {\tilde z} \partial z} ={\partial^2 e^\rho \over \partial t^2}.
~\rho =\rho (z,{\bar z},t). \eqno (16) $$ 
At this stage we should point out that one should $not$ confuse the variables $z,{\tilde z},t $ of eq-(16) with the previous $z,{\tilde z}$ coordinates and the ones to be discussed below. The operator form of the Moyal deformations of (16) 
may be obtained from the ( nontrivial) 
change of coordinates which takes $\Omega (w,{\tilde w},q,p;\hbar )$ to 
the function $u(t,{\tilde t},q',p';\hbar )$ 
defined as  :

 $$u(t,{\tilde t},q',p';\hbar ) \equiv \sum_{n=0}^{n=\infty} (\hbar)^n 
u_n (r'\equiv t{\tilde t};q',p'). \eqno (17)$$
The mapping of the effective $3D$ fields  
$\Omega_n (r\equiv w{\tilde w},q,p)$ appearing in the power expansion (7b) 
into the $ u_n (r'\equiv t{\tilde t};q',p')$, 
furnishes the Moyal deformed continuous Toda equation. 

The map of the zeroth-order  terms , $\Omega_o (r,q,p)\rightarrow u_o(r',q',p')$ is the analog of the map that [2,3] found to show how a rotational Killing symmetry reduction of the ( undeformed) Plebanski equation leads to the ordinary continuous Toda equation ( the first equation in the series appearing in (5)).  Roughly speaking, to zeroth-order,  having a function $\Omega (r,z,{\tilde z})$, one introduces a  new set of variables $t\equiv r\partial_r \Omega (r,z,{\tilde z})$; $s\equiv \partial_{{\tilde z}}\Omega$ and ${\tilde w} =z;w={\tilde z}$. After one eliminates $s$ and defines 
$r\equiv e^u$ , one gets the field $u=u(t,w,{\tilde w})$ which  satisfies  the continuous Toda equation, as a result of the $elimination$ of $s$, iff the 
original $\Omega (r,z,{\tilde z})$ obeyed the Killing symmetry reduction of Plebanski's equation to start with. The transformation from $\Omega$ to $u$ is a Legendre-like one.

Order by order in powers of $(\hbar)^n$
one can define :

$$t=t_o+\hbar t_1+\hbar^2 t_2....+\hbar^nt_n.~
t_n \equiv {r \partial \Omega_n (r,z,{\tilde z})\over \partial r}.~n=0,1,2,..$$

$$s=s_o+\hbar s_1+\hbar^2 s_2....+\hbar^ns_n.~
s_n \equiv {\partial \Omega_n (r,z,{\tilde z})\over \partial {\tilde z}}.~n=0,1,2,....\eqno ( 18)$$
this can be achieved  after one has solved iteratively eqs-(5) to order $n$  for every $\Omega_n (r,z,{\tilde z});$ with $ n=0,1,2...$.
After eliminating $s_0,s_1,s_2....s_n$ , to order $n$,  one has for analog of the original relation : $r=e^u$ the following :

$$r=r(t=t_o+\hbar t_1.....+\hbar^n t_n;z={\tilde w};{\tilde z}=w)\equiv e^{u_o +\hbar u_1 +...\hbar^n u_n}. \eqno (19)$$
eq-(19) should be viewed  as :

$$e^u =1 +(u_o +\hbar u_1....+\hbar^n u_n) +
{1\over 2!}(u_o +\hbar u_1....+\hbar^n u_n)^2+.......$$
$$r=r(t_0) +{\partial r \over \partial t}(t_o) (\hbar t_1....+\hbar^n t_n) +
{1\over 2!}{ \partial^2 r \over \partial t^2}(t_o) (\hbar t_1....+\hbar^n t_n)^2+.......$$
$$u_0=u_0(t_0;z,{\tilde z}).~u_1 =u_1 (t_0+\hbar t_1;z,{\tilde z}).....u_n =u_n (t_0+\hbar t_1+\hbar^2t_2.....+\hbar^n t_n;z,{\tilde z}).\eqno (20)$$
this procedure will allow us , order by order in powers of 
$(\hbar)^n$, after eliminating $s_o,s_1,s_2...$ to find the corresponding 
equations involving the functions $u_n (t,w,{\tilde w})$ iff the set of fields $\Omega_n$ obeyed eqs-(5) to begin with. It would be desirable  if one could have a master Legendre-like  transform from the function $\Omega(r,z,{\tilde z};\hbar)$ to the $u(t,w,{\tilde w};\hbar)$ that would generate all the equations in one stroke. i.e. to have a compact way of writing the analog of eqs-(2,5) for the field $u=\sum_n \hbar^n u_n$. In {\bf IV} we will define such transform.

A further dimensional reduction, $\Omega_n (r,z,{\tilde z})=\Omega_n (r;\tau=z+{\tilde z})$ for all $n$ yields the Moyal deformations of the $2D$ continuous Toda molecule : $u_n =u_n (t,\tau =w+{\tilde w})$.

 Going back to our original notation, by means of the dictionary relation (6), one can establish the maps and, in principle, recast/rewrite  the infinite number of equations (5) in terms of $u_n (r'=t{\tilde t},q',p')$.  
The operator analog amounts to relating the operator   
${\hat \Omega }(w,{\tilde w})$ with   the operator  ${\hat  u}(t,{\tilde t})$, consistent with the WWM transfom,  :

$$u(t,{\tilde t},q',p';\hbar )\equiv \int ^\infty_{-\infty}~d\xi <q'-{\xi\over 2}|{\hat  u}(t,{\tilde t})|q'+{\xi \over 2}> exp [{i\xi p'\over \hbar}].
\eqno (21)$$
where $u(t,{\tilde t},q',p')$ is given by (17).

Inverting gives :

$$  <q'-{\xi\over 2}|{\hat  u}(t,{\tilde t})|q'+{\xi \over 2}>\equiv \int ^\infty_{-\infty}~dp'~  u(t,{\tilde t},q',p';\hbar )exp [-{i\xi p'\over \hbar}].
\eqno (22)$$
the latter matrix elements suffice to determine the operator ${\hat u} (t,{\tilde t})$ associated with ${\hat \Omega}(w,{\tilde w})$ that satisfies the operator-valued zero curvature condition (14).

A further dimensional reduction  corresponds to  the deformed $2D$ continuous Toda molecule 
equation which should be equivalent to the Moyal deformations of the  $SU(\infty)$ Nahm's equations. The ansatz which furnished the map from the ordinary $SU(\infty)$ Nahm's equations to the $2D$ Toda  continous molecule in connection to the quantization of the self dual 
membrane was studied in [1]. As mentioned earlier, embedding the $SU(2)$ 
solutions found by [9] into the $SU(\infty)$ Moyal-deformed Nahm equations 
yields special solutions to the quantum $2D$ Toda molecule . A $U_\infty$ co-adjoint orbit action furnishes more.

To finalize, we may take an  alternative route.  The continuous Toda molecule equation as well as the usual Toda system  may be written in the double commutator form of the Brockett equation [18] :

$${\partial L (r,\tau) \over \partial r} =[L,[L,H]]. \eqno (23)$$
$L$ has the form

$$L\equiv A_{+} + A_{-} =X_o (-iu)+X_{+1} (e^{(\rho/2)})
+X_{-1} (e^{(\rho/2)}). \eqno (24)$$
with the connections $A_{\pm}$ taking values in the subspaces 
${\cal G}_o \oplus {\cal G}_{\pm 1}$ of some {\bf Z}-graded continuum Lie algebra ${\cal G}=\oplus _m {\cal G}_m $ of a novel class.  
$H=X_o (\kappa)$ is a continuous limit of the Cartan element of the principal 
$sl(2)$ subalgebra of ${\cal G}$. The functions $\kappa (\tau),u(r,\tau),\rho (r,\tau)$ 
satisfied certain equations given in [18].

One may consider the case when the group {\bf G} is the group of unitary operators acting in the Hilbert space of square integrable functions on the line, 
$L^2 (R^1)$. Then, ${\cal G}$ is now the ( continuum ) Lie algebra of 
self-adjoint   
operators acting in the Hilbert space, $L^2 (R^1)$. 
The operator-valued ( acting in the Hilbert space) quantities depending on the two coordinates, $r,\tau$ that obey 
the operator version of the 
Brockett equation and whose WWM map is :

 $${\partial {\hat L} (r,\tau) \over \partial r} =
 {1\over i\hbar}  [ {\hat L},  {1\over i\hbar} [ {\hat  L},{\hat H}]]. 
\leftrightarrow  {\partial { \cal L}  \over \partial r} =
   \{ { \cal   L},  \{ { \cal  L},{ \cal H}\}\}. \eqno (25)$$
where ${\cal L} (r,\tau,q,p;\hbar),{\cal H}(r,\tau,q,p;\hbar)$ are the corresponding elements in the phase space after  performing the WWM map. The main problem with this approach is that we do $not$ have representations of the continuum {\bf Z}-graded Lie algebras in the Hilbert space, $L^2(R^1)$ and , consequently, we cannot evaluate the matrix elements $<q-{\xi\over 2}|{\hat L}(r,\tau)|q+{\xi \over 2}>;<{\hat H}>$.  
For this reason we have to recur to the iterations in (5) and insert the solutions into (10a,10b).

In section {\bf IV} we will come back to eqs-(23-25) and show how the master Legendre-like transform between eqs-(2,5) and the continuum Toda theory can be achieved by using the Brockett equation.  
The supersymmetric extensions follow from the results of [4] where we wrote down the Plebanski analog of $4D$ Self Dual  Supergravity .

To conclude this section : A WWM formalism is very appropriate to Moyal quantize the continous Toda theory which we believe is the underlying theory behind the self dual membrane. 
Due to the variable entaglement of the original Toda equation, given by the first equation  in  the series of eqs-(5), one has to use the dictionary relation (6) that allows to use the WWM formalism of [9] in a straightforward fashion.

\bigskip

\centerline {\bf III. $SU_q(\infty)$ Moyal-Nahm Equations}

The solutions to the $SU(2)$ Moyal-Nahm equations found  by [9] 
:

$${\partial A^i \over \partial \tau} ={1\over 2}\epsilon_{ijk} \{A_j,A_k\}_{Moyal}. \eqno (26a)$$

in terms of Jacobi elliptic functions are :

$$A^1 =sn(\tau,k)[{i\over 2}p(q^2-1)-\hbar (\beta+{1\over 2})q].
~A^2 =dn(\tau,k)[{-1\over 2}p(q^2+1)-i\hbar (\beta+{1\over 2})q].\eqno (26b)$$
$$
A^3 =cn(\tau,k)[-ipq +\hbar (\beta+{1\over 2})].
$$
The $\tau$ derivatives of the Jacobi elliptic functions are :

$${\partial \over \partial \tau}sn(\tau,k)=cn(\tau,k)dn(\tau,k).~
{\partial \over \partial \tau}cn(\tau,k)=-sn(\tau,k)dn(\tau,k).$$
$${\partial \over \partial \tau}dn(\tau,k)=-k^2sn(\tau,k)cn(\tau,k).\eqno (26c)$$

In the $\hbar =0$ limit one recovers solutions to the classical $SU(\infty)$ Nahm equations. Eqs-(26) will be our starting point to find solutions to the
$SU_q(\infty)$ Nahm Equations.
\smallskip

\centerline {\bf 3.1 The $q$-Star product}
\smallskip
One can obtain a further deformation of the $SU(2)$ Moyal-Nahm equations by using the notion of $q$-deformed star products and $q$-Moyal brackets [25].  
In this case one loses the underlying associative character of the algebras. 
Therefore, non-trivial deformations of the Poisson bracket, $other$ than the 
$\hbar$ (Moyal) deformations can be obtained. In terms of the Jackson $q$-derivative :

$$D_z f(z)\equiv {f(z)-f(qz)\over (1-q)z}. \eqno (27)$$
one can construct the $q$-star products :

$$f*^q g\equiv \sum_{r=0}^\infty {(i\hbar)^r\over [r]!}f(p,q)[\overleftarrow{D}^r_p~exp (lnq~ \overleftarrow{\partial}_p~pq~ \overrightarrow{\partial}_q)~\overrightarrow{D}^r_q ]g(p,q) \eqno (28)$$
where one should $not$ confuse the phase space variable $q$ with the $q$ deformation parameter. The $q$-Moyal bracket is :

$$\{f,g\}_{q-M}\equiv {1\over i\hbar}(f*^q g-g*^q f). \eqno (29)$$

In the $q=1$ limit one gets the original Moyal bracket; whereas in 
the $\hbar \rightarrow 0$ limit one gets the $q$-Poisson bracket for the observables : 

$$f(q,p)=\sum_i f_i (p,q).~ g(q,p)=\sum_j g_j (p,q).~f_i =p^{m_i}q^{n_i}.~
g_j =p^{m_j}q^{n_j}.\eqno (30)$$
the $f_i,g_j$ are monomials in $p,q$ : of the form $p^mq^n$. 
The bracket is :
 $$\{f,g\}_{q-PB}\equiv \sum_{ij}
q^{\alpha(f_i,g_i)}[{D}_pf_i]~exp (lnq~ \overleftarrow{\partial}_p~pq~ \overrightarrow{\partial}_q)~[{D}_q g_j] -$$
$$ q^{\alpha(g_j,f_i)}[{D}_pg_j]~exp (lnq~ \overleftarrow{\partial}_p~pq~ \overrightarrow{\partial}_q)~[{D}_q f_i]. \eqno (31)$$
and the exponents in powers of $q$ are $\alpha (p^mq^n,p^kq^l)=nk$.

We have arrived at the main point of this section : Since the 
$\hbar =0$ limit of the $SU(2)$ Moyal-Nahm equations are the classical 
$SU(\infty)$ Nahm equations [9] , the $\hbar =0$ limit of the $SU(2)$ $q$-Moyal-Nahm equations corresponds  to the $SU(\infty)$ $q$-Nahm equations!  
Solutions of the $SU_q(\infty)$ Nahm equations  can be obtained as follows :

Define the $q$-deformed YM potentials as a  series expansion of the $q$ analog of spherical harmonics [26,27].  The  $\tau$ dependence may involve $q$-Jacobi elliptic functions [36] but for the time being we shall only concentrate on $q$ deformations of the spherical harmonics  :

$$A^1_q = {i\over 2}  [sn(\tau,k)] \sum_{JM}A^1_{JMq} \Psi^J_{Mq}.
~A^2_q = (-{1\over 2})    [dn(\tau,k)] \sum_{JM}A^2_{JMq} \Psi^J_{Mq}.$$
$$A^3_q =-i[cn_(\tau,k)] \sum_{JM}A^3_{JMq} \Psi^J_{Mq}. \eqno (32)$$
where the $q$ analog of the spherical harmonics are defined in terms of the $q$-Vilenkin functions [26,27]:

$$\Psi^J_{M0q}=C_{J0q}i^{-2J+M}{1\over \sqrt {[2J,q]!}}P^J_{M0q}(cos\theta)e^{-iM\phi}. \eqno (33)               $$
where $P^J_{M0q}(cos\theta)$ are the $q$-Vilenkin functions :
$$  P^J_{M0q}(cos\theta)=i^{2J-M}\sqrt {{[J+M,q]![J,q]!\over [J-M,q]![J,q]!
  }}[{1+cos\theta \over 1-cos\theta}]^{M/2} Q_{Jq}(\zeta)R^J_{Mq}(\zeta)            \eqno (34) $$
and the functions, $Q_{Jq}(\zeta);R^J_{Mq}(\zeta)$ are suitable functions of 
: $\zeta \equiv  (1+cos\theta)/( 1-cos\theta)$.

The $SU(2)$ $q$-Moyal-Nahm equations are, in the $\hbar =0$ limit, 
 the $SU_q(\infty)$ Nahm equations  :

$$D_q (\tau) A^i_q ={1\over 2}\epsilon^{ijk}\{A^j_q, A^k_q\}_{q-PB}
\eqno (35a)$$
where $D_q (\tau) A^i_q $ is  the Jackson derivative of $A^i_q$ w.r.t $\tau$ defined in (27).  This is 
where the $q$-Jacobi elliptic functions [36] should appear in (32) instead of the ordinary Jacobi elliptic functions. Without loss of generality, for the time being let us look at the equations :

$$ {\partial \over \partial \tau} A^i_q ={1\over 2}\epsilon^{ijk}\{A^j_q, A^k_q\}_{q-PB}
\eqno (35b)$$

The $q$-Poisson brackets amongst the $q$ spherical harmonics are :

$$\{\Psi_{JMq}, \Psi_{J'M'q}\}_{q-PB}= _qF^{M~M'~M+M'}_{ J~J'~J''} \Psi_{J''M''q}. \eqno (36)$$
where we have omitted the $0$ index in the definition of the spherical harmonics $\Psi _{JMq}$ in terms of the $q$-Vilenkin functions.  The stucture constants, $_qF^{m,m',m''}_{j,j',j''}$ are the $q$ extension of the classical structure constants of the area-preserving diffs of the sphere : $sdiff~S^2$.

It was  found by [6] that in the $N\rightarrow \infty$ limit, structure constants of the $sdiff~S^2$ coincide with the $SU(N)$ structure constants,  if a suitable basis was chosen. Such basis was assigned [6]  by selecting for each  $Y^m_j$ a symmetric traceless homogeneous polynomials in the variables $x,y,z$ subject to the constraint $x^2+y^2+z^2 =r^2$. $SU(N)$ emerges in the truncation of the spherical harmonics to a finite set by restricting $j\leq N-1$ so that the net sum of the number of generators, $Y^m_l$,  for $-j \leq m\leq j$ yields $N^2-1$ which is the number of independent generators of $SU(N)$. The correspondence was 
[6] :

$$Y^m_j \rightarrow T_{jm} =4\pi ({N^2-1 \over 4})^{(1-j)/2}~a^{jm}_{i_1i_2...
...i_j} L_{i_1}L_{i_2}.......L_{i_j}. \eqno (37)$$
and the angular momentum generators $L_i$ satisfy the equations :

$$[L_i, L_j]=i\epsilon_{ijk}L_k.~L^2 ={N^2-1\over 4}{\hat 1}. \eqno (38)$$

Therefore, a basis choice for the generators of $SU(N)$ are the $T_{jm}$ so that :

$$[T_{j_1m_1}, T_{j_2,m_2}]=i f_{j_1m_1, j_2m_2}^{j_3m_3}T_{j_3m_3}. 
\eqno (39)$$ 
In the $N\rightarrow \infty$ limit it was shown [6] that the structure constants $f_{j_1m_1, j_2m_2}^{j_3m_3}$ in eq-(39)  coincide with the $F^{M~M'~M+M'}_{J~J'~J''}$, of the area-preserving diffs of $S^2$. The latter are the  classical $q=1$ limit of the $q$ structure constants , 
$_qF^{M~M'~M+M'}_{J~J'~J''}$, in eq-(36).

The $q$ deformed case , $_qf_{j_1m_1, j_2m_2}^{j_3m_3} $,  requires to use the $q$ extension of the $3j$ and $6j$ symbols that appear in (39). Thus, the $N\rightarrow \infty$ limit of the quantity below yields the expression for the $q$ structure constants in eq-(36) : 

$$_qf^{M~M'~M+M'}_{J~J'~J''}=-4\pi i \prod_{i=1}^3 \sqrt {[2l_i +1]_q}
~[_qC^{j_1,j_2,j_3 }_{m_1,m_2,m_3 }][_qW^{j_1,j_2,j_3}_{s,s,s}]
[(-1)^N  {  R_{Nq} (j_1) R_{Nq}(j_2)\over R_{Nq} (j_3)}].\eqno (40)$$
where the $q$ analog of the $3j$ ($_qC^{...}_{....}$) and $6j$ 
($_qW^{....}_{...}$) symbols can be found in [35]. The value of $s$ was 
$2s=N-1$.  
The $q$ extension of the functions $R_N$ is defined as :

$$R_{Nq}(j)\equiv (N^2-1)^{(j-1)/2} \sqrt { { [N+j]_q!\over [N-j-1]_q!}}. 
\eqno (41)$$
where the $q$-integers and $q$ generalization of factorials are defined :

$$[n]_q\equiv {q^n -q^{-n} \over q-q^{-1}}.~~
[n]_q! \equiv [{q^n -q^{-n} \over q-q^{-1}}]! =[n][n-1].......[1].~~[0]_q!=1. 
\eqno (42)$$
The large $N$ limit of (40) should coincide  with the $q$ structure constants ,$_qF^{m,m',m''}_{j,j',j''}$ associated with the $q$ deformations of the area preserving diffs of the sphere. The structure constants only differ from zero 
iff :

$$ |j_1-j_2|+1\leq j_3 \leq j_1 +j_2 -1.~~m+m'+m''=0.~~ j_1+j_2+j_3 =odd. 
\eqno(43)$$

The $N\rightarrow \infty$ limit of : 

$$[_qW^{j_1,j_2,j_3}_{s,s,s}]
[(-1)^N {R_{Nq} (j_1) R_{Nq}(j_2)\over R_{Nq} (j_3)}].\eqno (44)$$
is :

$$(-1)^{j_3-1}[1+j_1+j_2+j_3]_q [j_1]_q![j_2]_q![j_3]_q!
\sqrt {{[j_1+j_2-j_3]_q![j_1+j_3-j_2]_q![j_2+j_3-j_1]_q! \over 
[1+j_1+j_2+j_3]_q!}}.$$
$$\sum_{n=0}^{j_1+j_2-j_3} (-1)^n [n]_q .
\{ [n]_q![j_1+j_2-j_3 -n]_q! [j_1-n]_q![j_2-n]_q!
[n+j_3-j_1]_q![n+j_3-j_2]_q!\}^{-1}. \eqno (45)$$

Plugging the expression (32) for the $SU_q(\infty)$ YM potentials   
into  (35b) yields the equations for the $q$-coefficients appearing in the expansion (32) after matching term by term multiplying each $\Psi^J_{Mq}$. It is much simpler, however,  to recur to the known solutions of the classical $SU(\infty)$ Nahm equations [9] and to establish the following correspondence :

$$A^i (\tau,q,p;\hbar=0 )=f^i(\tau,k)A^i_{jm}Y^m_j \rightarrow 
f^i(\tau,k)A^i_{jm} \Psi^j_{mq}=A^i_q (\tau,q,p). \eqno (46)$$
where the $q$ analog of the spherical harmonics is introduced in the r.h.s and where we maintain the ordinary Jacobi elliptic functions for the $\tau$ dependence.  Therefore, setting :

$$A^1_q =sn(\tau,k) \sum_{jm} A^i_{jm} \Psi^{J=j}_{M=m,q}(\theta,\phi).~
~A^2_q = dn(\tau,k)\sum_{jm}A^2_{jm} \Psi^{J=j}_{M=m,q}(\theta,\phi)$$
$$A^3_q = cn(\tau,k)\sum_{jm} A^3_{jm} \Psi^{J=j}_{M=m,q}(\theta,\phi). 
\eqno (47)$$
is a suitable ansatz for simple nontrivial solutions to the $SU_q(\infty)$ Nahm equations, (35b).  The coefficients, $A^i_{jm}$, can be obtained as follows :

Firstly,  one must have the the map from the $p,q$ variables to the $cos~\theta,\phi$ ones is performed via the stereographic projection of the sphere to the complex plane:
This is required bacause the  $(q,p)$ variables live in the two-dim plane.

$$z=q+ip =\rho e^{i\phi}.~{\bar z}=q-ip =\rho e^{-i\phi}.~q=\rho~ cos~\phi.~
p=\rho~ sin~\phi. ~\rho=cot~{\theta\over 2}.\eqno (48)$$ 
in this way functions on the sphere, $f(\theta,\phi)$ can be projected onto functions $f(z,{\bar z})=f(q,p)$. Poisson brackets are now taken w.r.t the $cos\theta,\phi$ variables. The same applies for $q$ derivatives.   

Secondly, the $\hbar =0$ limit of (26) yields the solutions to the classical $SU(\infty)$ nahm equations [9]  :

$$A^1 ={i\over 2}sn(\tau,k)[p(q^2-1)].~A^2 =-{1\over 2}dn(\tau,k)[p(q^2+1)].~A^3 =(-i) cn(\tau,k)[pq]. \eqno (49)$$

Hence, the  coefficients, $A^i_{jm}$ in (47)  can be obtained directly from eqs-(49) after using the orthonormality property of the spherical harmonics :

$$
A^1_{jm}={i\over 2}\int\int p(q^2-1) [Y^{m}_j]^* d\phi~ d(-cos\theta).
A^2_{jm}=-{1\over 2}\int \int p(q^2+1) [Y^{m}_j]^* 
d\phi~ d(-cos\theta).         \eqno (50)$$

$$A^3_{jm}=-i\int^{2\pi}_0 \int^1_{-1} pq [Y^{m}_j]^* d\phi~ d(-cos\theta).\eqno (51)$$

The mapping from $(q,p)$ to $(\theta,\phi)$ thus allows us to compute the coefficients (50,51) in terms of trigonometric integrals after using the general formula :

$$Y^m_l ={(-1)^l\over 2^l l!}\sqrt {{ (2l+1)(l+m)!\over 4\pi (l-m)!}} e^{im\phi}
(sin~\theta)^{-m} {d^{l-m}\over d(cos~\theta)^{l-m}}(sin~\theta)^{2l}.~
[Y^m_l]^*=(-1)^m Y^{-m}_l. \eqno (52)$$
and the recurrence relations :

$$(cos~\theta) Y^m_l=\sqrt{  {(l+m+1)(l-m+1)\over (2l+1)(2l+3)}}Y^m_{l+1}+
\sqrt{  {(l+m)(l-m)\over (2l+1)(2l-1)}}Y^m_{l-1}. \eqno (53)$$

To conclude this section , eqs-(47) with coefficients given by (50,51) $are$ the simplest 
nontrivial   solutions to the $SU_q(\infty)$ Nahm equations (35b)  based on the $\hbar =0$ limit of the solutions to the  $SU_q(2) $ Moyal-Nahm equations.
It is straighforward to verify that in the $q\rightarrow 1$ limit, the $SU_q(\infty)$ YM 
potentials (47) become the classical $SU(\infty)$ YM ones given by eqs-(49).
Also, the $q$ structure constants become the classical ones . The same occurs with  
the  $SU_q(\infty)$ Nahm equations : these  become the classical 
Nahm equations.   
Solutions to the original equations, (35a), where the Jackson time derivative is used,  $must$ involve the use of $q$ Jacobi elliptic functions.

\bigskip

\centerline{\bf 3.2 The $SU_q(\infty)$ Nahm Equations and the $SU_q(\infty)$ Toda molecule}
\smallskip

Now we shall proceed to 
embed the $SU_q(N)$ Toda molecule into the $SU_q(\infty)$ Nahm equations. 
In [1] we have shown that the classical continuous Toda molecule can be embedded into the $SU(\infty)$ classical Nahm equations by a suitable ansatz. The latter equations are equivalent to the lightcone-gauge self dual ( spherical) membrane equations of motion ( moving in flat target spacetime backgrounds). We will follow exactly the same steps here by simply extending the ordinary Poisson bracket to the $q$ Poisson case and similarily with  all the other quantities. 
The authors [2] have also discussed the $SU(2)$ Toda molecule in connection to the self dual membrane in $5D$. The lightcone gauge is an effective $4D$ $SU(\infty)$ SDYM theory dimensionally reduced to one temporal dimension : the 
$SU(\infty)$ Nahm equations, in the temporal gauge $A_0=0$. 

Reductions of the $SU(2)$ Moyal-Nahm equations in connection to the continuous Toda molecule, in the $\hbar =0$ limit, have also been discussed by [16]. The reduction required an ansatz which allows one to resum exactly the infinite series in the Moyal bracket. One recovers the classical continuous Toda molecule in the $\hbar =0$ limit. This reduction procedure is precisely the one we shall use in the next section in order to embed the $2D$ Moyal-Toda 
( continuous) molecule into the $SU(\infty)$ Moyal-Nahm equations; i.e. into Moyal deformations of the self dual membrane.  

Essential in the formulation of the $SU_q(\infty)$ Toda molecule is the notion of $Quantum$ Lie algebras [30]  associated with the quantized universal enveloping algebras. An example is the universal  $U_q (su(\infty)$ algebra. This  will be discussed next in connection with the $q$ deformations of the $SU(N)$ Toda chain. The $quantum$  Lie algebra associated with the universal enveloping area-preserving diffs algebra, $U_q (su(\infty)$, is 
denoted by ${\cal L}_q [su(\infty)]$ [30].  
The symmetry algebras associated with the $q$ deformations of the Toda theory are the $q$ $W_\infty $ type algebras. The construction of $q$-Virasoro and $q$- $W_\infty$ algebras has been discussed by [32,33]; a readable discussion of the $q$-$W$ KP hierarchies of quantum integrable systems has been given by [34].

In this section we shall embed the $q$-deformed continuous Toda molecule
into the $SU_q(\infty)$ Nahm equations in a straightforward fashion. To start with, it is 
relevant to show how one relates the continuuum limit of $\Theta^q_J (\tau)$ 
to the continuous Toda field : $\rho^q(\tau,t)$.

The mapping from the discrete $\Theta^q_j (\tau)$ in the continuum limit to the continuous $\rho^q (\tau,t)$ field is :

$$t_j =t_o +j\epsilon.~\epsilon ={t_f-t_o \over N}\sim 
{\rho^q (\tau,t_f)-\rho^q(\tau,t_o)\over  \Theta^q_N(\tau)-\Theta^q_0(\tau)}. \eqno (54)$$
 
$$\epsilon [\Theta^q_j (\tau)-\Theta^q_0 (\tau)]=
[\rho^q(\tau,t_j)-\rho^q(\tau,t_0)].~lim~{N\rightarrow \infty;\epsilon \rightarrow 0}~:N\epsilon =t_f-t_0.  \eqno (55)$$
 
when $\epsilon \rightarrow 0$ the continuum field  $\rho^q (\tau, t_f)
= \rho^q(\tau,t_0)\rightarrow 0$ 
for nonzero $\Theta^q_0,\Theta^q_N$. These values of $\rho^q (\tau, t_f)
= \rho^q(\tau,t_0)\rightarrow 0$ are consistent with the trivial solutions to the continuum Toda molecule equation. The Cartan matrix is :

$$K^q (t,t')=\epsilon K_{jj'}.~t_0 +j\epsilon \leq t\leq t_0 +(j+1)\epsilon.~
  t_0 +j'\epsilon \leq t'\leq t_0 +(j'+1)\epsilon.\eqno ( 55b)$$

Once solutions to the $\Theta^q_J (\tau)$ are known, eqs-(54,55) yield the continuum limit : $\rho^q (\tau,t_j)\equiv \epsilon \Theta^q_j (\tau)$.

Before one can embed the $q$ continuous Toda molecule equations into the $SU_q(\infty)$ Nahm equations we must  have a precise understanding of what one means by a $q$ $SU(N)$ Toda theory and by a quantum root system.   
A natural generalization of ordinary Lie algebras was introduced in [30] . 
These algebras were coined $quantum$ Lie algebras. 
The $quantum$ Lie algebras, ${\cal L}_q (g)$, are certain adjoint submodules of quantized universal enveloping algebras $U_q (g)$. They are non-associative algebras which are embedded in the quantized enveloping algebras, $U_q(g)$ of Drinfeld and Jimbo in the same way as ordinary Lie algebras are embedded into their enveloping algebras. The $quantum$ Lie algebras  are endowed with a $quantum$ Lie bracket given by the quantum adjoint action : $[AoB]=AoB$ . The structure constants and the quantum root system are now functions of $q$ and they exhibit duality symmetry under $q\leftrightarrow (1/q)$.

The $q$ deformed algebra ${\cal L}_q (g)$ has the same dimension as $g$ and is invariant under the so called tilded Cartan involutions, ${\tilde S}$, tilded antipode action, 
${\tilde \theta}$ and automorphisms, $\tau$,  of the diagrams. The $quantum$ ``Cartan subalgebra'' spanned by the generators $H_i$ have $non$-vanishing quantum Lie brackets among themselves and there are $two$ sets of quantum roots. For $sl_n$ , these roots form a $quantum$ root $lattice$ : if $a_\alpha, a_\beta$ are two roots then $a_\alpha + a_\beta =a_{{\alpha +\beta}}$ and $a_{-\alpha} =-a_\alpha$.  However, this is $not$ true for other algebras, like $sp(4)$, for example.  A quantum analog of the symmetric bilinear Killing form exists which is invariant under the quantum adjoint action. For $sl(n)$ the analog of the Cartan matrix is :

$$K^q_{ij}\equiv K(H_i, H_j)=b [ (q+q^{-1})\delta_{ij} -\delta_{i,j-1} -\delta_{i,j+1}]. \eqno (56a)$$
with  $b(q)$ a $q$ dependent normalization constant. In the particular
case of the $sl_3$ algebra, the authors [30] found :
$$b =[(q^{-1/2}+q^{1/2})^2 (q^{-3/2}+q^{3/2})^2]{C^2 \over 4}. \eqno (56b)$$
with $C$ a $q$-dependent normalization factor

$$\{ 2(q^{-1/2} +q^{1/2})(q^{-3/2} +q^{3/2})(q^{-3}+q^{-1}-1+q+q^3)\}^{-1/2}. \eqno (56c) $$
See [30] for further details.   When $q=1$ one recovers the standard Cartan matrix for $sl(3)/su(3)$, up to a multiplicative factor.

  The $quantum$ Lie algebra ,$ {\cal L}_q (su(\infty) $  is the one associated with the universal enveloping algebra, $U_q(su(\infty)$. For example, $U_q(sl_n)$ is an algebra over the ring of formal power series in $q$,  and can be defined [28,29] in terms of the Cartan elements, $h_i$, the raising elements, $e_i$, and the simple lowering elements $f_i$ that obey the relations :

$$[h_i, h_j]=0.~[h_i,e_j]=A_{ij}e_j.~[h_i,f_j]=-A_{ij}f_j.~[e_i,f_j]=\delta_{ij}{q^{h_i}- q^{-h_i}\over q -q^{-1}}. \eqno (57)$$
plus the quadratic and cubic Serre relations :

$$[e_i,e_j]=0.~[f_i,f_j]=0~|i-j|\ge 2. \eqno (58a)$$

$$e^2_i e_j -(q+q^{-1})e_ie_je_i +e_je^2_i =0.~|i-j|=1.\eqno (59b)$$ 

$$f^2_i f_j -(q+q^{-1})f_if_jf_i +f_jf^2_i =0.~|i-j|=1. \eqno (59c)$$ 
where $A_{ij}$ is the Cartan matrix, $K_{jj'}$, for $su(N)$.  
The Hopf algebra structure of $U_q(su(n)$ is given by a comultiplication, antipode and counit, respectively : $\Delta, S, \epsilon$ . 

$$\Delta (h_i)=h_i\otimes 1 +1\otimes h_i.~
\Delta (e_i)=e_i\otimes q^{-h_i/2} + q^{h_i/2}\otimes e_i......\eqno (59d)$$
$$S(h_i)= -h_i.~ S(e_i)=-q^{-1}e_i.~S(f_i)=-q^{1}f_i.~\epsilon (h_i)=
\epsilon (e_i)=\epsilon (f_i)=0. \eqno (59e)$$
We refer to [28,29] for more details.  With these definitions, now one can have a precise meaning of 
what one means by the    
 $SU_q(N)$ Toda chain comprised of the $q$-deformed fields : $\Theta^q_J (\tau)$,  that take values in the $quantum$ Cartan subalgebra of the $quantum$ Lie algebra
${\cal L}_g (su(\infty)$.   

The $q$-analog of the ansatz in [1] requires to choose the special class of solutions to the $SU_q(\infty)$ Nahm equations using 
the definitions : $A^q_y \equiv A^q_1 +i A^q_2$ and $ A^q_{{\tilde y}} \equiv A^q_1 -i A^q_2$. The $q$ analog of the ansatz in [1] reads :

$$A^q_y =\sum_j A^q_{y,j} (\tau) \psi^{+1}_{q j} (\theta,\phi). ~A^q_{\tilde y} =\sum_{j} 
A^q_{{\tilde y},j} (\tau)\psi^{-1}_{jq} (\theta,\phi).
~A^q_3=\sum_{j} A^q_{3,j} (\tau)\psi^{0}_{jq} (\theta,\phi).\eqno (60)$$
where now the functions $\psi_{jmq} (\theta,\phi)$ belong to a certain subspace of the $\Psi_{JMq} (\theta,\phi)$ $q$-spherical harmonics. Representations of the quantum group ,$SU_q(2)$, associated with the $U_q (su(2)$ enveloping algebra,  are constructed in terms of the $q$ spherical harmonics, $\Psi_{JMq}$,  and, correspondingly, representations of the $quantum$ algebra,  
${\cal L}_q (su(2)$ algebra, associated with the universal enveloping algebra, 
$U_q(su(2))$, can be constructed in terms of $\psi_{jmq}$. Similarily, the $q$-Poisson bracket defined in (31) for the functions $\Psi_{JMq}$,  $induces$ a ${\cal L}_q (PB)$ bracket on the functions $\psi_{jmq} (\theta,\phi)$.

The $\tau$ dependence of the coefficients in (60) is again taken to be  :

$$A^q_{1,j}(\tau)\equiv {i\over 2}sn(\tau,k)a^q_{1,j}.~ 
A^q_{2,j}(\tau) \equiv -{1\over 2}dn(\tau,k) a^q_{2,j}    .~  
A^q_{3,j}(\tau) \equiv  (-i)cn(\tau,k)a^q_{3,j}.\eqno (61)$$
where one could  include the $q$-Jacobi elliptic functions if one were to use the Jackson derivative w.r.t $\tau$. 

The $q$-extension of the ansatz in [1] requires then that the  
${\cal L}_q (PB)$-Poisson bracket of $A^q_y, A^q_{{\tilde y}}$ corresponds to    :

$$\{A^q_y, A^q_{\tilde y}\}_{{\cal L}_q(PB)}=\sum_{j} exp[K^q_{jj'}\Theta^q_{j'}(\tau)]
\psi_j^{0}. \eqno (62a)$$
$${\partial A^q_z \over \partial \tau}=\sum_{j} {\partial^2 \Theta^q_j (\tau)\over \partial \tau^2}\psi_j^{0}. \eqno (62b)$$
where $K^q_{jj''}$ is the $quantum$ Cartan matrix (56) and the $\Theta^q_j (\tau)$ are the $q$ deformed Toda fields associated with $SU_q(N)$ Toda chain/molecule.    Upon evaluating  the ${\cal L}_q$-Poisson brackets between the 
functions , $\psi_j^{+1}, \psi_{j'}^{-1}$,  in the l.h.s of (62a), yields terms involving the  $\psi_{j''}^{0}$ in the r.h.s. 
This is consistent with the quantum Lie bracket relations of the ${\cal L}_q (sl_2)$ [30]. 
Matching term by term multiplying each  function, $\psi^0_{jq}$ yields the system of equations relating special solutions of the $SU_q(\infty)$ Nahm equations and the $SU_q(\infty)$ Toda.

It is important to emphasize that the solutions of the 
$q$ deformed continuous Toda chain which are being embedded into the special class of solutions of the $SU_q(\infty)$ Nahm equations (60)  do $not$ 
correpond  with the earlier solutions of the Nahm equations found in eqs-(47)! This one can see by inspection. The functions $\psi_{jmq}$ are not the same as $\Psi_{JMq}$; the Poisson brackets and the YM potentials are not either. 
Therefore, eqs-(62) determine the link between  the $\Theta^q_J (\tau)$ with 
 the special class of solutions of the $SU_q(\infty)$ Nahm equations in the 
$SU_q(N=\infty)$ limit. In particular from (62), due to the orthogonality of the $q$ spherical harmonics  one learns that :

$$\Theta^q_j (\tau)=
\int^\tau _0 d\tau ' \int A^q_z (\theta,\phi,\tau')[\psi^0_j]^* (\theta,\phi) [d\Omega]_q. \eqno (62c)$$
where a $q$ measure of integration [27]  , the $q$ solid angle, must be used in the r.h.s of (62c).
Therefore, solutions of the type (62c) can be linked to solutions of the 
$q$ deformed $SU_q(N)$ Toda chain in the $N\rightarrow \infty$ limit. 
Once the $\Theta^q_J (\tau)$ are found , eqs-(54,55) determine the values of the continuous $q$-Toda  field, $\rho^q (\tau,t)$.  

We finalize this section by writing down the 
$q$ deformation of the continuous Toda chain/molecule equations.
The continuum limit of the quantum Cartan matrix given by (56a)  is :

$$K^q (t,t')=b[(q+q^{-1} -2)\delta (t-t') +\partial^2_t~\delta (t-t')]. \eqno 
(63a)$$
where we simply added and subtracted, $2\delta_{ij}$ from (56a); 
and the $SU_q(\infty)$ Toda molecule  equation  is :

$${\partial^2 \over \partial \tau^2}\rho_q (\tau,t) =b(q)[(q+q^{-1}-2)+
{\partial^2 \over \partial t^2}]~e^{\rho_q (\tau,t)}. \eqno (63b) $$
where  $b(q)$ is a normalization factor.   

Rigorously speaking, we must have the Jackson derivative in the l.h.s of (63b)  so that the true $q$ continuum Toda molecule reads :

$$D_q^2 (\tau)\rho_q (\tau,t) =b(q)[(q+q^{-1}-2)+
{\partial^2 \over \partial t^2}]~e^{\rho_q (\tau,t)}. \eqno (63c) $$
and $q$-Jacobi elliptic functions for the $\tau$ dependence of the YM potentials must be included.

The relevance of studying these $q$ Toda models is due to their  exact quantum integrability. One can calculate exact quantum masses in terms of the quantum twisted Coxeter numbers, exact location of the poles of the $S$ matrices, three point couplings,...[30,43.44,45]. The large $N\rightarrow \infty$ limit will furnish an integrable sector of the membrane's mass spectrum associated with the quantum deformations of the self dual spherical membrane in flat target backgrounds. This is achieved via the correspondence with the $SU(\infty)$ Nahm equations. Since the continuum Toda theories have the  $W_\infty$ algebras as symmetry algebras, highest weight irreducible 
representations of the $W_\infty$,$q$-$W_\infty$ algebras will provide important tools in  the classification of the spectrum [1].

\bigskip

\centerline {\bf IV. The $SU(\infty)$ Moyal-Nahm equations}
\smallskip

We will study in this section the Moyal-Nahm equations. To begin with, the Moyal bracket of the YM potentials appearing in equations like (35) , $A_y, A_{{\tilde y}}$, can be expanded in powers of $\hbar$  as  [16] :

$$\sum_{s=0}^\infty {(-1)^2\hbar^{2s}\over (2s+1)!}
\sum_{l=0}^{2s+1} (-1)^l (C^{2s+1}_l )[\partial_q^{2s+1-l}\partial_p^l A_y]
[\partial_p^{2s+1-l}\partial_q^l A_{\tilde y}]. \eqno (64)$$
where $C^{2s+1}_l$ are the binomial coefficients.

The crucial difference between the 
solutions of the $SU(2)$ Moyal-Nahm eqs [9] and the $SU(\infty)$ Moyal-Nahm 
case is that one $must$ have an $extra$ explicit dependence on another variable, $t$, for the YM potentials.  For example, expanding  in powers of $\hbar$ , the YM potentials involved in the $SU(\infty)$ Moyal-Nahm equations are :

$$A^i(\tau,t,q,p;\hbar)\equiv \sum_{n=0}^\infty \hbar^n~A^i_n (\tau,t,q,p). 
\eqno (65) $$

In this fashion, in the continuum limit, $N\rightarrow \infty$, we expect to have the continuous Moyal-Toda molecule and be able to embed it  into the $SU(\infty)$ Moyal-Nahm equations and, hence, the relationship to the Moyal deformations of self dual membrane will be established. Similar results hold in the supersymmetric case. It was for this reason that the highest weight representations of $W_\infty$ algebras should provide important information about the self dual membrane spectra [1]. Dimensional reduction of $W_\infty\oplus {\bar W}_\infty$ act as the spectrum generating  algebra.

Matters in the Moyal deformations are no longer that simple because of the infinite number of derivative terms appearing  in the Moyal bracket. If one had the analog of the $q$-spherical harmonics one could sum the infinite series and reorganize the terms accordingly as one did in eqs-(36).  
For this reason we shall discuss briefly the results of [16] where the 
Moyal-Nahm equations admit a reduction to the Toda chain. 

The $SU(2)$ Moyal-Nahm equations are of the form : 

$${\partial X  \over \partial \tau }= \{Y,Z\}.~{\partial Y \over \partial \tau }= \{Z,X\}.
~{\partial Z  \over \partial \tau }= \{X,Y\}. \eqno (66)$$
The ansatz :

$$X =h(q,\tau)cos~p.~X =h(q,\tau)sin~p. ~Z=f(q,\tau). \eqno (67)$$
allowed [16] to resum the infinite series in the Moyal bracket. After the field redefinition $e^{\rho/2} =h(q,\tau)$ , one obtains  the Toda chain equation upon elimination of the function $f(q,\tau)$ :

$${\partial^2 \rho \over \partial \tau^2}=-
[{\Delta -\Delta^{-1} \over \hbar}]^2 e^{\rho}.~\rho =\rho (q,\tau). \eqno (68)$$
The operators in the r.h.s of (68)  are defined [16] as the shift operators : $\Delta f =f(q+\hbar).~\Delta^{-1}f  =f(q-\hbar).$ In the classical $\hbar =0$ limit one recovers the continuum Toda chain. The operator term in the r.h.s , when $\hbar =0$, is exactly the continuum limit of the Cartan $SU(N)$ matrix : $K(q,q')=\ \partial_q ^2 \delta ( q-q')$ [18]. 
This can be easily seen by writing the Cartan matrix  :

$$K_{ij}\equiv -(\delta_{i,j+2}-\delta_{ii}) -(\delta_{i,j-2}-\delta_{ii})
\rightarrow {-\Delta^{+2}+2-\Delta^{-2} \over \hbar^2}=-[{\Delta^1 -\Delta^{-1} \over \hbar}]^2.  \eqno (69)$$

From section {\bf II} one knows how to obtain ( in principle, by iterations ) solutions to the Moyal deformations of the effective $2D$ Toda molecule 
 equations , starting from solutions to the Moyal deformations of the rotational Killing symmetry reductions of Plebanski first heavenly equation. This is attained after one has performed the Legendre-like transform from the $\Omega_n (r, \tau=z+{\tilde z})$ fields  to the 
$u_n(t,\tau=w+ {\tilde w})$ for $n=0,1,2....$.  
Based on eqs-(68),  can one write an ansatz which 
encompasses the Legendre-like transform of the infinite number of eqs-(2,5)  
into one compact single equation involving the Moyal-deformed field :
$\rho (q,\tau;\hbar)$ ?. 

The clue was  provided earlier in {\bf II} by writing down the Toda equations in the double-commutator Brockett form [18]  given in eqs-(23-25)  : 

First of all one must have :

$$A_1 (t,\tau,p,q;\hbar) =\sum_{n=0}^\infty \hbar^n A^n_1(t,\tau,p,q).~
A_{2} (t,\tau,p,q;\hbar) =\sum_{n=0}^\infty 
 \hbar^n A^n_2,(t,\tau,p,q)
.\eqno (70)$$

$$A_{3} (t,\tau,p,q;\hbar) =\sum_{n=0}^\infty 
 \hbar^n A^n_3(t,\tau,p,q)
.\eqno (71)$$

Secondly, one may impose the correspondence with eqs-(23-25). Given  
${\cal L} (\tau,t,q,p;\hbar)$ and ${\cal H} (\tau,t,q,p;\hbar)$ :

$$ \{A_1, \{A_3  , A_1 \} \}_{Moyal}+ \{ \{A_2, A_3 \}, A_2  \}_{Moyal} 
\leftrightarrow \{ {\cal L}  , \{ {\cal L}, 
{\cal H} \} \}_{Moyal}. \eqno (72)$$

$$ {\partial^2 A_3 (q,p,\tau,t;\hbar)  \over \partial \tau ^2} \leftrightarrow 
{\partial \over \partial \tau} {\cal L} (\tau,t,q,p;\hbar). \eqno (73)$$

In the l.h.s of (72,73), we rewrote the $SU(\infty)$ Moyal-Nahm eqs, as :

$${\partial^2 A_3   \over \partial \tau ^2}=
\{A_1, \{A_3  , A_1 \} \}_{Moyal}+ \{ \{A_2, A_3 \}, A_2  \}_{Moyal}
.\eqno (74) $$

Eqs-(74) are obtained by a $straight$ dimensional reduction of the 
original $4D~SU(\infty) $ SDYM equations, which is an effective $6D$ theory, to the final effective equations in $4D$. The temporal gauge condition $A_0=0$ is required.  Whereas the r.h.s of (72,73) are  obtained through a sequence of reductions from the effective $6D$ theory  $\rightarrow 4D~SDG$  $\rightarrow 3D$ continuous Toda and, finally, to the continuous $2D$ Toda molecule.  The gauge conditions $A_y=A_z=0$ are required ( see [5])  and a WWM formalism is performed in order to recover quantities depending on the phase space variables, $p,q$.  
As mentioned earlier, to evaluate explicitly  the r.h.s of (72,73) requires a knowledge of representations,  in the Hilbert space $L^2 (R)$, of the {\bf Z} graded continuum Lie algebras described in [18]. As far as we know these have not been constructed.  

Therefore, concluding, the candidate master Legendre-like transform that takes the original Moyal-Plebanski equation $\{ \Omega_w, \Omega_{{\tilde w}} \}_M =1$ for all fields, $\Omega_n;~n=0,1,2...$, after the Killing symmetry and dimensional reductions,  into the Moyal-Toda equations $is$ the one given by eqs-(72,73). The number of variables matches exactly. The only difficulty is to write down representations of continnum {\bf Z}-graded Lie algebras in $L^2(R)$.  

When the Legendre-like transform equations are recast in terms of the 
$\rho (\tau', q';\hbar)$  field  one has , then,  the  compact single equation encompassig the Moyal deformations of the continuous Toda chain/molecule, after expanding in powers of $\hbar$ :   

$$ \rho (q',\tau';\hbar)\equiv \sum_{n=0}^\infty \hbar^n \rho_n (q',\tau '). \eqno (75)$$

A plausible guess, based on (68,69), and on eq-(63c),   is to write :

$$ {\cal D}^{2}(\tau';\hbar) \rho (q',\tau ';\hbar)=
- \sum_{n=0} \hbar^n a_n [{\Delta -\Delta^{-1} \over \hbar}]^{n+2}~ e^{\rho (q',\tau ';\hbar)}. \eqno (76)$$
with $a_n$ being coefficients,  as a tentative Moyal deformation of the continuum Toda molecule. The derivative ${\cal D}(\tau;\hbar) $ is some deformation of the ordinary derivative
w.r.t the $\tau'$ variable. For example, one could use the Jackson derivative for the $q$ parameter $q=e^\hbar$.  
This does $not$ mean that (76) is the correct equation ! The correct Legendre-like transform $is$ the one provided by eqs-(72,73) in terms of the Brockett double commutator form.  

The real test to verify whether or not (76) , indeed, is the correct continuous Moyal-Toda  chain equation   is to study  the Legendre-like transform of the infinite number of eqs-(2,5) and see whether or not it agrees with eqs-(76). One could  use eqs-(76) as the $defining$ master Legendre-transform. The question still remains if such transform is compatible and consistent with eqs-(2,5) and with eqs-(18-20). And, furthermore, whether or not it agrees with eqs-(72,73) as well! Clearly one has to integrate eqs-(72,73) w.r.t the $q,p$ variables in order to have a proper match of variables with those appearing in eqs-(76).  This  difficult question is currently under investigation. 

To conclude this section, eqs-(72,73) encode the 
master Legendre-like transform between the  Moyal-Plebanski equations and the $SU(\infty)$ Moyal-Nahm equations associated with the Moyal-Toda molecule.

\centerline{\bf V. Concluding Remarks}.

We hope to have advanced the need to recur to $q$-Moyal deformation quantization methods applied to the light-cone gauge self dual spherical membrane 
moving in flat target backgrounds : a $SU(\infty)$ $q$-Moyal-Nahm system.  The full membrane, for arbitrary topology,  moving in curved backgrounds remains to be studied. The $q$ deformations of the continuum Toda theory associated with the $q$-Moyal-Nahm system must admit a spectrum of the same type like the solitons of quantum affine Toda theories. This was already noticed by the author in [1] pertaining to the spectrum of noncritical ( linear and nonlinear) $W_\infty$ strings. A BRST analysis revealed that $D=27, D=11$ were the target spacetime dimensions , if the theory was devoid of quantum anomalies, for the bosonic 
  and supersymmetric case , respectively. These are also the 
 target dimensions of the ( super) membrane. This seems to indicate that there is a noncritical $W_\infty$ string sector  within  the self dual membrane.   
This sector is the one related to the affine continuum Toda theory.

The study of the quantum affine Toda theory allows, among many other things,   the exact calculation of full quantum mass ratios [43,44,45]  in terms of quantum twisted Coxeter numbers  associated with the quantum Lie algebras , ${\cal L}_q (g)$  discussed by [30]. In particular for $g=su(\infty)$.   
It is for this main physical reason that a $q$-Moyal deformation quantization 
program may be very helpful in understanding important features of the quantum theory of membranes, once we are able to extend the domain of validity beyond the integrable self dual case. This is where the representation theory of $W_\infty$ and 
$q$-$W_\infty$ algebras will be an essential tool. 

Deformation quantization techniques have been discussed by [37]. The geometric quantization of the $q$ deformation of current algebras living in $2D$ compact Riemann surfaces has been discussed by [38]. 
The role of noncommutative geometry [39] and the   quantum differential geometry of the quantum phase space associated with the Moyal quantization has been analysed by 
[31,40]. The role of matrix models in the membrane quantization was studied in 
[41].  The importance of integrability in the Seiberg-Witten
theory and strings has been studied by [42]. Deformations of the Kdv hierarchy 
and related soliton equations have been previuosly analyzed in [46] and the relations between Affine Toda solitons with Calogero-Sutherland  and spin
Ruijsenaars-Schneider models have been studied in [47].

To finalize, we wish to point out that there seems to be $two$ levels of quantization , one level is the Moyal 
quantization and the other level is the quantum group method. A combination of both Moyal and quantum group deformations should yield  a ``third'' level of quantization which was not 
explored here.  

\centerline {Acknowledgements}
\smallskip

We thank  I.A.B Strachan for sending us the proof of how the  Moyal-Nahm equations admit reductions to the continuum Toda chain. 

\smallskip

\centerline {\bf REFERENCES}

1-C. Castro :'' On the Integrability aspects of the Self Dual Membrane ``

hep-th/9612241. Phys. Lett {\bf B 288} (1992) 291.

E.G. Floratos, G.K. Leontaris : Phys. Lett. {\bf B 223} (1989) 153.

2- C.D. Boyer, J.Finley III : Jour. Math. Phys. {\bf 23} (6) (1982) 1126. 

C.Boyer, J.Finley, J.Plebanski : `` Complex Relativity and ${\cal H}$,

${\cal H} {\cal H}$ spaces `` General Relativity and Gravitation, Einstein 

Memorial volume. A Held ( Plenum 1980) vol. 2 pp 241-281.

3-Q.H. Park : Int. Jour. Mod. Physics {\bf A 7} (1991) 1415.

4- C. Castro : J. Math. Phys. {\bf 34} (1993) 681.

Jour. Math. Phys {\bf 35} no.6 (1994) 3013. 

5-J. Plebanski, M. Przanowski : Phys. Lett {\bf A 219} (1996) 249.

6- The original proof was given by : J.Hoppe : `` Quantum Theory of a Relativis

tic Surface `` M.I.T Ph.D thesis  (1982).  

B. de Wit, J. Hoppe, H. Nicolai : Nucl. Phys. {\bf B 305} (1988) 545.

B.Biran, E.G. Floratos, G.K. Saviddy : Phys. Lett {\bf B 198} (1987) 32. 

7- G. Chapline, K. Yamagishi : Class. Quan. Grav. {\bf 8} (1991) 427.  

8-E. Nissimov, S. Pacheva : Theor. Math. Phys. {\bf 93} (1992) 274.

9- H. Garcia-Compean, J. Plebanski : `` On the Weyl-Wigner-Moyal description 

of $SU(\infty)$ Nahm equations'' hep-th/9612221.

10-D. Diaconescu `` $D$ branes, Monopoles and Nahm equations `` hep-th/9608163.

11-J.S. Park : `` Monads and $D$ Instantons `` hep-th/9612096

12- A. Hanany, E.Witten : `` $IIB$ Superstrings , BPS Monopoles and 

$3D$ Gauge Dynamics `` hep-th/9611230.

13-Y. Hashimoto, Y. Yasui, S.Miyagi ,T.Otsuka :``Applications of Ashtekar 

gravity to $4D$ hyper-Kahler Geometry and YM instantons.``hep-th/9610069.

14-A.Ashtekar, T. Jacobson, L. Smolin : Comm. Math. Phys. {\bf 115} (1988) 

631.

15-C. Castro : Phys. Lett. {\bf B 353} (1995) 201.

16-I.A.B  Strachan : Phys. Lett {\bf B 282} (1992) 63. ``The Geometry of 

Multidimensional Integrable systems'' hep-th/9604142. `` The dispersive 

self-dual Einstein equations and the Toda lattice''. Private communication.  

17-H. Weyl : Z. Phys. {\bf 46} (1927) 1.

E. Wigner : Phys. Rev. {\bf 40} (1932) 749.

J. Moyal : Proc. Cam. Phil. Soc. {\bf 45} (1945) 99. 

18-M. Saveliev : Theor. Math. Phys. {\bf 92} (1992) 457.

M. Saveliev, A. Vershik : Phys. Lett {\bf A 143} (1990) 121.

19- D.B. Fairlie, J. Nuyts : Comm. Math. Phys. {\bf 134} (1990) 413.

20- I. Bakas, B. Khesin, E. Kiritsis : Comm. Math. Phys. {\bf 151} (1993) 233.

21-C. Pope, L. Romans, X. Shen : Phys. Lett. {\bf B 236} (1990) 173.

22- K.B. Wolf : `` Integral Transform Representations of $SL(2,R)$  in Group 

Theoretical Methods in Physics `` K.B. Wolf ( Springer Verlag, 1980) 

pp. 526-531

23-C.L.  Metha, E.C.G. Sudarshan : Phys. Rev. {\bf 138 B} (1963) 274. 

24- M.J. Ablowitz, P.A. Clarkson : ``Solitons, Nonlinear Evolution Equations 

and Inverse Scattering `` London Math. Soc. Lecture Notes vol {\bf 149} 

Cambridge Univ. Press, 1991. R.S Ward : Phys. Lett {\bf A 61} (1977) 81. 

Class. Quan. Grav. {\bf 7} (1990) L 217.  

25-O.F Dayi : ``q-deformed Star Products and Moyal Brackets `` q-alg/9609023. 

I.M Gelfand, D.B Fairlie : Comm. Math. Phys {\bf 136} (1991) 487. 

26-G. Rideau, P. Winternitz : Jour. Math. Phys. {\bf 34} (12) (1993) 3062.

27-J.R. Schmidt : Jour. Math. Phys. {\bf 37} (6) (1996) 3062.

28-V.G. Drinfeld : Sov. Math. Doke {\bf 32} (1985) 254.

29-M. Jimbo : Lett. Math. Phys. {\bf 10} (1985) 63.

30- G.W. Delius, A. Huffmann : `` On Quantum Lie Algebras and Quantum Root 

Systems ``q-alg/9506017.

G.W. Delius, A. Huffmann, M.D. Gould, Y.Z. Zhang  : `` Quantum Lie Algebras 

associated with $U_q(gl_n)$ and $U_q (sl_n)$'' q-alg/9508013.

V. Lyubashenko, A. Sudbery : `` Quantum Lie Algebras of the type $A_n$ `` 

q-alg/9510004.

31-M.Reuter : ``Non Commutative Geometry on Quantum Phase Space ``

hep-th/9510011. 

32-E.H. El Kinani, M. Zakkari : Phys. Lett. {\bf B 357} (1995) 105. 

J. Shiraishi, H. Kubo, H. Awata, S. Odake : `` A quantum deformation of the 

Virasoro algebra and Mc Donald symmetric functions `` q-alg/9507034 

33- C. Zha : Jour. Math. Phys. {\bf 35} (1) (1994) 517. 

E. Frenkel, N. Reshetikhin : `` Quantum Affine Algebras and Deformations of the 
Virasoro and $W$ algebras `` : q-alg/9505025. 

34- J. Mas, M. Seco : Jour. Math. Phys. {\bf 37} (12) (1996) 6510. 

35- L.C. Biedenharn, M.A. Lohe : `` Quantum Group Symmetry and $q$-Tensor

Algebras `` World Scientific, Singapore, 1995. Chapter 3. 

36- R. Floreanini, J. le Tourneaux, L. Vincent :  Jour. Math. Phys. {\bf 37}

(8) (1996) 4135.

37-G.Dito, M. Flato, D. Sternheimer, L. Takhtajan : ``Deformation quantization 

of Nambu Mechanics `` hep-th/9602016.

38- S. Albeverio,  S.M. Fei : `` Current algebraic structures over manifolds ,

Poisson algebras, $q$-deformation Quantization `` hep-th/9603114.

39- A. Connes : ``Non commutative Differential Geometry `` Publ. Math. IHES 

62 (1985) 41. 

40- B.V. Fedosov : J. Diff. Geometry {\bf 40} (1994) 213.  

H. Garcia Compean, J. Plebanski, M. Przanowski : ``Geometry associated with 

self-dual Yang-Mills and the chiral model approaches to self dual gravity ``

hep-th/ 9702046    

41- A. Jevicki : `` Matrix models, Open strings and Quantization of 

Membranes'' hep-th/9607187. 

T. Banks, W. Fischler, S.H. Shenker, L. Susskind `` M theory as 

a Matrix Model, a conjecture ``hep-th/9610043.

42- A. Marshakov : `` Nonperturbative Quantum Theories and Integrable Equations

`` ITEP-TH-47/96 preprint ( Lebedev Phys. Institute) 

H. Itoyama, A. Morozov : `` Integrability and Seiberg-Witten Theory'' 

ITEP-M 7- 95 / OU-HET-232 preprint. 

43-H.W. Braden, E. Corrigan, P.E. Dorey, R. Sasaki : Nuc. Phys {\bf B 338} 

(1990) 689.

L. Bonora, V. Bonservizi : Nucl. Phys. {\bf B 390} (1993) 205. 

P. Christie, G. Musardo : Nucl. Phys. {\bf B 330} (1990) 465. 

D. Olive, N. Turok, J. Underwood :  Nucl. Phys. {\bf B 409} (1993) 509.

T. Hollowood : Nucl. Phys. {\bf B 384} (1992) 523.

44- G.W. Delius, M.T Grisaru, D. Zanon :  Nuc. Phys {\bf B 382 } (1992 ) 365 .

45-P.E Dorey :  Nuc. Phys {\bf B 358 } (1991 ) 654 .

46- E. Frenkel : `` Deformations of the Kdv Hierarchy and related Soliton 

equations `` q-alg/9511003

47- H. Braden, A. Hone : `` Affine Toda solitons of the Calogero-Moser type ``

hep-th/ 9603178.

I. Krichever, A. Zabrodin : `` Spin generalization of the Ruijsenaars-Schneider 
model, nonabelian $2D$ Toda chain and representations of the Sklyanin algebra ``

hep-th/9505039.

\end